\documentclass[hyper]{JHEP} % 10pt is ignored!

\usepackage{epsfig}
\newcommand\fverb{\setbox\pippobox=\hbox\bgroup\verb}

\newcommand\fverbdo{\egroup\medskip\noindent%

            \fbox{\unhbox\pippobox}\ }

\newcommand\fverbit{\egroup\item[\fbox{\unhbox\pippobox}]}

\newbox\pippobox

\title{Note About Hamiltonian Formalism for General Non-Linear
Massive Gravity Action in  St\"{u}ckelberg Formalism}
\author{J. Kluso\v{n}\\
Department of
Theoretical Physics and Astrophysics\\
Faculty of Science, Masaryk University\\
Kotl\'{a}\v{r}sk\'{a} 2, 611 37, Brno\\
Czech Republic\\
E-mail: \email{klu@physics.muni.cz}}
\preprint{}

 \abstract{In this note we try to prove the absence of the ghosts in case of the
 general non-linear massive gravity action in St\"{u}ckelberg
 formalism. We argue that in order to find the explicit form of the
 Hamiltonian it is natural to start with the general non-linear
 massive gravity action found in
arXiv:1106.3344 [hep-th].  We
 perform the complete Hamiltonian analysis of
the St\"{u}ckelberg form of
 the minimal
 the non-linear gravity action
in this formulation and show that the constraint structure is so
rich that it is possible to eliminate non-physical modes. Then we
extend this analysis to the case of the general non-linear massive
gravity action. We find the corresponding Hamiltonian and collection
of the primary constraints. Unfortunately we are not able to finish
the complete analysis of the stability of all constraints due to the
complex form of one primary constraint so that we are not able to
determine the conditions under which  given constraint is preserved
during the time evolution of the system.} 
\def\tSigma{\tilde{\Sigma}}
\def\mM{\mathcal{M}}

\def\mR{\mathcal{R}}
\def\tK{\tilde{K}}
\def\tn{\tilde{n}}
\def\bA{\mathbf{A}}

\def\tx{\tilde{x}}

\def\mC{\mathcal{C}}
\def\be{\begin{equation}}

\def\ee{\end{equation}}

\def\bea{\begin{eqnarray}}

\def\eea{\end{eqnarray}}

\def\tr{\mathrm{tr}\, }

\def\tr{\mathrm{Tr}}

\def\bx{\mathbf{x}}
\def\by{\mathbf{y}}

\newcommand{\hg}{\hat{g}}

\newcommand{\tD}{\tilde{D}}
\newcommand{\mG}{\mathcal{G}}

\def \bA{\mathbf{A}}

\newcommand{\bT}{\mathbf{T}}

\def\pb #1{\left\{#1\right\}}

\begin{document}
%%%%%%%%%%%%%%%%%%%%%
%%%%Introduction %%%%%%%%%
%%%%%%%%%%%%%%%%%%%%
\section{Introduction and Summary}
The first formulation of the massive gravity was performed by
\cite{Fierz:1939ix} at least in its linearized level as the
propagation of the massive graviton above the flat background
\footnote{For recent review and extensive list of references, see
\cite{Hinterbichler:2011tt}.}. Even if the theory seems to be well
defined at the linearized level there is a Boulware-Deser ghost
\cite{Boulware:1974sr} in the naive non-linear extension of the
Fierz-Pauli formulation.
 On the other hand recently there has been
much progress in the non-linear formulation of the massive gravity
without the Boulware-Deser ghost \cite{deRham:2010ik,deRham:2010kj}
and also \cite{Hassan:2011vm,Hassan:2011hr}\footnote{For related
works, see
\cite{Hassan:2012gz,Hassan:2012wr,Chiang:2012vh,Nojiri:2012zu,
Parisi:2012cg,Cai:2012ag,Nomura:2012xr,Saridakis:2012jy,Fasiello:2012rw,
D'Amico:2012pi,deRham:2012kf,Berg:2012kn,Gumrukcuoglu:2012aa,DeFelice:2012mx,
Kobayashi:2012fz,Baccetti:2012bk,Hassan:2012wt,Hinterbichler:2012cn,
Paulos:2012xe,Crisostomi:2012db,Golovnev:2011aa,Gumrukcuoglu:2011zh,
Berezhiani:2011mt,Comelli:2011zm,vonStrauss:2011mq,Comelli:2011wq,
deRoany:2011rk,Gumrukcuoglu:2011ew,D'Amico:2011jj,deRham:2011qq,deRham:2011rn,Huang:2012pe}.}.

The Hamiltonian treatment of non-linear massive gravity  theory was
performed in many papers with emphasis on the general proof of the
absence of the ghosts in given theory. The first attempt for the
analysis of the constraint structure of the non-linear massive
gravity  was performed in \cite{Kluson:2011qe}. However it turned
out that this analysis  was not complete and the wrong conclusion
was reached as was then shown in the fundamental paper
\cite{Hassan:2011ea} where the complete Hamiltonian analysis of the
gauge fixed  form of the general non-linear massive gravity was
performed. The fundamental result of given paper is the proof of the
existence of two additional constraints in the theory which  are
crucial  for the elimination of  non physical modes and hence for
the consistency of the  non-linear massive gravity at least at the
classical level.

Then the Hamiltonian analysis of the non-linear massive gravity in
the  St\"{u}ckelberg  formulation  was performed in
\cite{Kluson:2011rt,Kluson:2012gz}. Unfortunately  the wrong
conclusion was again reached in the first versions of given papers
as was then shown in \cite{Hassan:2012qv} where  the absence of the
ghosts in the minimal version of non-linear massive gravity was
proven for the first time. Then an independent proof of the absence
of the ghosts in the minimal version of non-linear massive gravity
in the St\"{u}ckelberg formulation
 was presented
in  \cite{Kluson:2012wf}.

However the proof of the absence of the ghosts in the general form
of the non-linear massive gravity in the St\"{u}ckelberg formulation
is still lacking. The difficulty with the possible Hamiltonian
formulation of given theory is that the action depends on the
kinetic terms of the St\"{u}ckelberg fields in a highly non-linear
way so that it seems to be impossible  to find an explicit  relation
between canonical conjugate momenta and the time derivatives of the
St\"{u}ckelberg fields. On the other hand there exists the
formulation of the non-linear massive gravity action with the linear
dependence on the  kinetic term for the St\"{u}ckelberg fields. This
is the form of the non-linear massive gravity action that arises
from the original one when the redefinition of the  shift functions
is performed
\cite{Hassan:2011hr,Hassan:2011vm,Hassan:2011zd,Hassan:2011tf}. The
goal of this paper is to perform the Hamiltonian analysis of given
action and try to identify all constraints.

As we argued above the  main advantage of
 the formulation of the non-linear massive gravity with redefined
 shift function  is that
the kinetic term for the St\"{u}ckelberg fields appears linearly and
hence it is easy
 to find the corresponding Hamiltonian even for the general form
of the non-linear massive gravity action. Further it is possible to
identify  four primary constraints of the theory where the three
ones are parts of the generators of the spatial diffeomorphism. Note
that the presence of the diffeomorphism constraints is the
reflection of the fact that we have manifestly diffeomorphism
invariant theory. On the other hand the  fourth primary constraint
could be responsible for the elimination of the additional non
physical mode.
 However this claim is only true when the requirement
of the preservation of given constraint during the time development
of the system generates another additional constraint. Unfortunately
we find that the original primary constraint cannot provide such
additional constraint due to the fact that the Poisson bracket
between the primary constraints defined at different space points is
non zero. For that reason we should find another constraint that
obeys the property that Poisson bracket between these constraints
defined at different space points is zero.  We find such a
constraint in the case of the minimal non-linear massive gravity
action and we show  that this constraint has the same form as the
primary constraint found in \cite{Kluson:2012wf}. Then we will be
able to show that the requirement of the preservation of given
constraint during the time evolution of the system implies the
additional constraint and these two constraints together allow to
eliminate two non-physical modes. This result agrees with previous
two independent analysis performed in \cite{Hassan:2012qv} and
 in \cite{Kluson:2012wf}.

Unfortunately we are not able to reach the main goal of this paper
which is the proof of the absence of the ghosts for the general
non-linear massive gravity theory in St\"{u}ckelberg formalism. The
reason is that we are not able to find the primary constraint that
has vanishing Poisson bracket between these constraints defined at
different space points and that has the Poisson brackets with
another constraints that vanish on the constraint surface. This is
crucial condition for the existence of the additional constraint.
 It is rather worrying  that we are not able to finish the
Hamiltonian analysis for the general non-linear massive gravity
action especially in the light of the very nice proof of the absence
of the ghosts  in case of the gauge fixed non-linear massive gravity
action \cite{Hassan:2011ea}. However there is a possibility that the
proof of the absence of the ghosts for general non-linear massive
gravity action in St\"{u}ckelberg formalism could be found in the
very elegant formulation of the massive and multi metric theories of
gravity presented in \cite{Hinterbichler:2012cn}. We hope to return
to this problem in future.

The structure of this paper is as follows. In the next section
(\ref{second}) we introduce the non-linear massive gravity action
and perform the field redefinition of the shift function. We also
 we find the Hamiltonian formulation of
the minimal form of the non-lineal massive gravity in
St\"{u}ckelberg formulation with redefined shift functions and we
find that given theory is free from the ghosts. Then we extend this
approach to the case of the general non-linear massive gravity
theory in the section (\ref{third}). We identify all primary
constraints and  discuss the difficulties that prevent us to finish
the complete Hamiltonian analysis.
%%%%%%%%%%%%%%%%%%%%%%%%%%%%%%%%%%%%%%%%%55
\section{Non-linear Massive Gravity with Redefined
Shift Functions}\label{second} As we stressed in the introduction
section the goal of this paper is to perform the Hamiltonian
analysis of the general non-linear massive gravity with presence of
the St\"{u}ckelberg fields. It turns out that it is useful to
consider this action with redefined shift functions
\cite{Hassan:2011hr,Hassan:2011vm,Hassan:2011zd,Hassan:2011tf}. More
explicitly, let us begin with following general form of the
non-linear massive gravity action
\begin{equation}
S=M_p^2 \int d^4x\sqrt{-\hg} [{}^{(4)}R+2m^2\sum_{n=0}^3 \beta_n
e_n(\sqrt{\hg^{-1}f})] \ ,
\end{equation}
where $e_k(\bA)$ are elementary
symmetric polynomials of the
eigenvalues of $\bA$. For generic
$4\times 4$ matrix they are given by
\begin{eqnarray}
e_0(\bA)&=&1 \ , \nonumber \\
e_1(\bA)&=&[\bA] \ , \nonumber \\
e_2(\bA)&=&\frac{1}{2}([\bA]^2-[\bA^2]) \
, \nonumber \\
e_3(\bA)&=&\frac{1}{6} \left( [\bA]^3-3[\bA][\bA^2]+2[\bA^3]\right)
\
, \nonumber \\
e_4(\bA)&=&\frac{1}{24} \left([\bA]^4-6[\bA]^2[\bA^2]+3
[\bA^2]^2+8[\bA][\bA^3]-6[\bA^4]\right)
\ , \nonumber \\
e_k(\bA)&=&0 \ , \  \mathrm{for} \  k>4 \ , \nonumber \\
\end{eqnarray}
where $\bA^\mu_{ \ \nu}$ is $4\times 4$
matrix and where
\begin{equation}
[\bA]=\tr \bA^\mu_{\ \mu} \ .
\end{equation}
Of the four $\beta_n$ two combinations are related to the mass and
the cosmological constant while the remaining two combinations are
free parameters. If we consider the case when the cosmological
constant is zero and the parameter $m$ is mass, the four $\beta_n$
are parameterized in terms of the $\alpha_3$ and $\alpha_4$ as
\cite{deRham:2010kj}
\begin{equation}
\beta_n=(-1)^n\left(\frac{1}{2}(4-n)(3-n)-
(4-n)\alpha_3+\alpha_4\right) \ .
\end{equation}
The minimal action corresponds to
$\beta_2=\beta_3=0$ that implies
$\alpha_3=\alpha_4=1$ and consequently
$\beta_0=3 \ , \beta_1=-1$.

We consider the massive gravity with  that is manifestly
diffeomorphism invariant. This can be ensured  with the help of $4$
scalar fields $\phi^A \ , A=0,1,2,3$ so that
\begin{equation}
\hg^{\mu\nu}f_{\nu\rho}=\hg^{\mu\nu}
\partial_\nu\phi^A\partial_\rho\phi_A \ .
\end{equation}
Then we have
\begin{equation}
N^2\hg^{-1}f= \left(\begin{array}{cc}
-f_{00}+N^l f_{l0} & -f_{0j}+N^l f_{lj}
\\
N^2 g^{il}f_{l0}-N^i(-f_{00}+N^l f_{l0}) & N^2
g^{il}f_{lj}-N^i(-f_{0j}+ N^l f_{lj}) \\ \end{array}\right) \ ,
\end{equation}
where we also used $3+1$ decomposition of the four dimensional
metric $\hat{g}_{\mu\nu}$ \cite{Gourgoulhon:2007ue,Arnowitt:1962hi}
\begin{eqnarray}
\hat{g}_{00}=-N^2+N_i g^{ij}N_j \ , \quad \hat{g}_{0i}=N_i \ , \quad
\hat{g}_{ij}=g_{ij} \ ,
\nonumber \\
\hat{g}^{00}=-\frac{1}{N^2} \ , \quad \hat{g}^{0i}=\frac{N^i}{N^2} \
, \quad \hat{g}^{ij}=g^{ij}-\frac{N^i N^j}{N^2} \ .
\nonumber \\
\end{eqnarray}
Let us now perform the redefinition of the shift function $N^i$ that
was introduced in
\cite{Hassan:2011vm,Hassan:2011hr,Hassan:2011zd,Hassan:2011tf}
\begin{equation}\label{defNi}
N^i=M\tn^i+f^{ik}f_{0k}+N \tD^i_{ \ j } \tn^j \ ,
\end{equation}
where
\begin{equation}\label{deftx}
 \tx=1-\tn^i f_{ij}\tn^j \ , \quad
 M^2=-f_{00}+f_{0k}f^{kl}f_{l0}
\end{equation}
%\begin{equation}
%N^i=c^i+N d^i \ ,
%\end{equation}
%where
%\begin{eqnarray}
%c^l&=&M^2 n^l+f^{ik}f_{k0} \ , \quad d^i=D^i_{ \ k}
%[c^k-f^{kl}f_{l0}]
% \ , \nonumber \\
% M^2&=&-f_{00}+f_{0k}f^{kl}f_{l0} \ , \nonumber \\
% \end{eqnarray}
and where we defined $f^{ij}$ as the inverse to $f_{ij}$ in the
sense \footnote{Note that in our convention $f^{ik}$ coincides with
$({}^{3}f^{-1})^{ik}$ presented in
\cite{Hassan:2011zd,Hassan:2011tf,Hassan:2011hr,Hassan:2011vm}.}
\begin{equation}
f_{ik}f^{kj}=\delta_i^{ \ j} \ .
\end{equation}
Finally note
%that  Inserting (\ref{deftilde}) into (\ref{Deq}) we
%find
that the matrix
 $\tD^i_{ \ j}$ obeys the equation
%\begin{equation}
%d^i=D^i_{ \ j}n^j=
%\tD^i_{ \ j} \tn^j \ ,
%\end{equation}
%where
%\begin{equation}
%n^k=c^k-({}^3 f^{-1})^{kl}f_{l0}=
%M \tn^k \
%\end{equation}
%and hence
\begin{eqnarray}
%\sqrt{x}D^i_{ \ j}=
%\sqrt{(g^{ik}-d^i d^k)f_{kl}} \Rightarrow \nonumber \\
\sqrt{\tx}\tD^i_{ \ j}=
\sqrt{(g^{ik}-\tD^i_{ \ m} \tn^m
\tD^k_{ \ n}\tn^n)f_{kj}} \  \nonumber \\
\end{eqnarray}
and also following important identity
\begin{eqnarray}
f_{ik}\tD^k_{ \ j}=
%f_{ik}\frac{1}{\sqrt{x}}\sum
%(c_0\delta^k_{ \ j}+ c_1 (Sf)^k_{ \
%j}+c_2 (SfSf)^k_{ \ j}+\dots)=
%\nonumber
%\\
%f_{jk}\frac{1}{\sqrt{x}}\sum
%(\delta^k_{ \ i}+ c_1
%S^{kl}f_{li}+\dots)=
f_{jk}
\tD^k_{ \ i } \ .  \nonumber \\
\end{eqnarray}

Let us now concentrate on the minimal form of the non-linear massive
gravity action.  Using the redefinition (\ref{defNi}) we  find that
it takes the form
\begin{eqnarray}\label{massgr2}
%S=M_p^2\int d^3\bx dt [N\sqrt{g}
%\tK_{ij}\mG^{ijkl}\tK_{kl}+{}^{(3)}R
%-2m^2(\sqrt{g}\tr\bA+N\sqrt{g}\tr
%\bB-3N\sqrt{g})]=
%\nonumber \\
S=M_p^2\int d^3\bx dt [N\sqrt{g}
\tK_{ij}\mG^{ijkl}\tK_{kl}+N\sqrt{g}R -\sqrt{g}MU
-2m^2(N\sqrt{g}\sqrt{\tx}D^i_{ \ i}-3N\sqrt{g})] \ ,
\nonumber \\
\end{eqnarray}
where
\begin{equation}
U=2m^2\sqrt{\tx}
 \ ,
 \end{equation}
 and
  where we used the $3+1$ decomposition of the four dimensional
scalar curvature
\begin{equation}\label{31R}
{}^{(4)}R=\tK_{ij}\mG^{ijkl}\tK_{kl}+R \ ,
\end{equation}
where $R$ is three dimensional scalar curvature and where
\begin{equation}
\mG^{ijkl}=\frac{1}{2}(g^{ik}g^{jl}+g^{il}g^{jk})- g^{ij}g^{kl}
\end{equation}
with inverse
\begin{equation}
\mG_{ijkl}=\frac{1}{2}(g_{ik}g_{jl}+g_{il}g_{jk})-\frac{1}{2}g_{ij}g_{kl}
\ , \quad
\mG_{ijkl}\mG^{klmn}=\frac{1}{2}(\delta_i^m\delta_j^n+\delta_i^n\delta_j^m)
\ .
\end{equation}
Note that  in (\ref{31R}) we  ignored the terms containing total
derivatives. Finally note that $\tK_{ij}$ is defined as
\begin{equation}
\tK_{ij}=\frac{1}{2N}(\partial_t g_{ij}- \nabla_i
N_j(\tn,g)-\nabla_j N_i(\tn,g)) \ ,
\end{equation}
where $N_i$ depends on $\tn^i$ and $g$ through the relation
(\ref{defNi}).

 At this
point we should stress the reason why we consider the non-linear
massive gravity action in the form (\ref{massgr2}). The reason is
that we want to perform the Hamiltonian analysis for the general
non-linear massive gravity action written in the St\"{u}ckelberg
formalism. It turns out that the action (\ref{massgr2}) has formally
the same form  as in case of the general non-linear massive gravity
action when we replace $U$ with more general form whose explicit
form was determined in
\cite{Hassan:2011vm,Hassan:2011hr,Hassan:2011zd,Hassan:2011tf}. On
the other hand the main advantage of the  action (\ref{massgr2}) is
that it depends on the time derivatives of $\phi^A$ through the term
$M$  and that this term  appears linearly in the action
(\ref{massgr2}). We should compare this fact with  the original form
of the non-linear massive gravity action where the dependence on the
time derivatives of $\phi^A$ is highly non-linear and hence it is
very difficult to find  corresponding Hamiltonian.

Explicitly, from (\ref{massgr2}) we find the momenta conjugate to
$N,\tn^i$ and $g_{ij}$
\begin{equation}
\pi_N\approx 0 \ , \pi_i\approx 0 \ ,
\pi^{ij}=M_p^2\sqrt{g}\mG^{ijkl}\tK_{kl} \
\end{equation}
and the momentum conjugate to $\phi^A$
\begin{eqnarray}
p_A
%2M_p^2\sqrt{g}\frac{\delta N_i}{\delta
%\partial_t\phi^A}
%\nabla_j\mG^{ijkl}\tK_{kl}-M_p^2\sqrt{g}\frac{\delta M}{\delta
%\partial_t\phi^A}U=\nonumber \\
=- \left(\frac{\delta M}{\delta\partial_t \phi^A}
\tn^i+f^{ij}\partial_j\phi_A\right) \mR_i-M_p^2\sqrt{g} \frac{\delta
M}{\partial_t \phi^A}
U \ , \nonumber \\
\end{eqnarray}
where
\begin{equation}
%U=2m^2\sqrt{\tx} \
 \mR_i=-2g_{ik}\nabla_j\pi^{kj} \ .
\end{equation}
It turns out that it is useful to write $M^2$ in the form
\begin{eqnarray}
M^2=-\partial_t\phi^A \mM_{AB}\partial_t \phi^B \ , \quad
\mM_{AB}=\eta_{AB}-\partial_i\phi_A f^{ij}\partial_j\phi_B \ ,
\nonumber \\
\end{eqnarray}
where by definition the matrix $\mM_{AB}$ obeys following relations
\begin{eqnarray}\label{mMprop}
\mM_{AB}\eta^{BC}\mM_{CD}=
% (\eta_{AB}-\partial_i\phi_A
%({}^{3}f^{-1})^{ij}\partial_j\phi_B) \eta^{BC}(\eta_{CD}-
%\partial_m\phi_D
%({}^{3}f^{-1})^{mn}\partial_n\phi_D)=
% \nonumber \\
% (\delta_A^{ \ C}-\partial_i\phi_A
%({}^{3}f^{-1})^{ij}\partial_j\phi^C) (\eta_{CD}-
%\partial_m\phi_D
%({}^{3}f^{-1})^{mn}\partial_n\phi_D)=
%\nonumber \\
%(\eta_{AD}-2\partial_i\phi_A({}^{(3)} f^{-1})^{ij}\partial_j\phi_D+
%+\partial_i\phi_A ({}^{(3)}f^{-1})^{ij} f_{jk}({}^{(3)}f^{-1})^{km}
%\partial_m\phi_D=
%\nonumber \\
%(\eta_{AD}-\partial_i\phi_A({}^{(3)} f^{-1})^{ij}\partial_j\phi_D=
\mM_{AD} \ , \quad  \det \mM^A_{ \ B}=1
\  \nonumber \\
\end{eqnarray}
together with
\begin{eqnarray}\label{partiphimM}
\partial_i\phi^A \mM_{AB}=
\partial_i\phi_B-
\partial_i\phi^A\partial_k\phi_A
f^{kl}\partial_l\phi_B=0 \ .
%\nonumber \\
%\partial_i\phi_A-f_{ik}({}^3 f^{-1})^{kl}
%\partial_l\phi_B=\partial_i\phi_A-
%\delta_i^{ \ l}\partial_l \phi_B=
%0 \nonumber \\
\end{eqnarray}
%\begin{eqnarray}
%\frac{\delta M}{\delta
%\partial_t\phi^A}=
%-\frac{1}{M}\mM_{AB}\partial_0\phi^B \
%. \nonumber \\
%%\frac{2}{2M}(-\partial_t \phi_A+\partial_i\phi_A ({}^3 f^{-1})^{ik}
%%f_{k0})=
%%\frac{1}{M}(-\eta_{AB}+\partial_i\phi_A ({}^3 f^{-1})^{ik}\partial_k
%%\phi_B)\partial_t\phi^B=\nonumber
%%\\
%%-\frac{1}{M}\eta_{AC}(\delta^C_{ \ B}-\partial_i\phi^C ({}^3
%%f^{-1})^{ik}\partial_k\phi_B)\partial_t\phi^B=
%%-\frac{1}{M}\eta_{AC}\mM^C_{ \ B}\partial_t\phi^B
%%\nonumber \\
%\end{eqnarray}
%An important property of the matrix $\mM^A_{ \ C}$
%is that it is a projector
%\begin{equation}
%\mM^A_{ \ B}\mM^B_{ \ C}=\mM^A_{ \ C} \
%\end{equation}
%and
%\begin{eqnarray}
%\det \mM^A_{ \ B}=
%\exp \tr \ln (\delta^A_{ \ B}-
%\partial_i\phi^C ({}^3
%f^{-1})^{ik}\partial_k\phi_B)=
%\nonumber \\
%=\exp  \sum_n c_n \tr
%(\partial_i\phi^C ({}^3
%f^{-1})^{ik}\partial_k\phi_B)^n=
%\exp (3c_0+c_1 f_{ik}({}^3
%f^{-1})^{ki}+c_2(f_{ik}({}^3
%f^{-1})^{kj}f_{jl}({}^3
%f^{-1})^{li})+\dots)=
%\nonumber \\
%=\exp (3c_0+3c_1+3c_2\dots)=
%\exp 3 \ln 1=\exp 0=1
%\nonumber \\
%\end{eqnarray}
%\begin{eqnarray}
%\partial_i\phi^A \mM_A^{ \ B}=
%\partial_i\phi^B-
%\partial_i\phi^A\partial_k\phi_A
%({}^3 f^{-1})^{kl}\partial_l\phi^B=
%\nonumber \\
%\partial_i\phi^A-f_{ik}({}^3 f^{-1})^{kl}
%\partial_l\phi^B=\partial_i\phi^A-
%\delta_i^{ \ l}\partial_l \phi^B=
%0 \nonumber \\
%\end{eqnarray}
With the help of these results we find
\begin{eqnarray}\label{PiAM}
%\Pi_A=p_A+ g_{ik}({}^3 f^{-1})^{kl}\partial_l \phi_A \mR^i= -
%(\tn_i\mR^i+M_p^2\sqrt{g}U) \frac{\delta M}{\delta\partial_t\phi^A}
%\Rightarrow
%\nonumber \\
p_A+\mR_if^{ij}\partial_j\phi_A= (\tn^i\mR_i+M_p^2\sqrt{g}U)
\frac{1}{M}\mM_{AB}\partial_t \phi^B \
%, \mM_{AB}\partial_t \phi^B=\frac{M}{
%(\tn_i\mR^i+M_p^2\mV)}\Pi_A \ .
%\nonumber \\
\end{eqnarray}
%where
%\begin{equation}
%\Pi_A=p_A+ g_{ik}f^{kl}\partial_l \phi_A \mR^i \ .
%\end{equation}
and consequently % Then with the help of (\ref{mMprop}) and
%(\ref{PiAM}) we find
\begin{eqnarray}
M^2=-\partial_t\phi^A\mM_{AB}\partial_t\phi^B=
%-\partial_0\phi^A\mM_{AC}\eta^{CD}\mM_{DB}\partial_0\phi^B=
%\nonumber \\
-\frac{M^2}{(\tn^i
\mR_i+M_p^2\sqrt{g}U)^2}(p_A+\mR_if^{ij}\partial_j\phi_A)
\eta^{AB}(p_B+\mR_i f^{ij}\partial_j\phi_B)
\nonumber \\
\end{eqnarray}
which however implies following primary constraint
\begin{equation}\label{defSigmap}
 \Sigma_p=(\tn^i
\mR_i+M_p^2\sqrt{g}U)^2+(p_A+\mR_if^{ij}\partial_j\phi_A) (p^A+\mR_i
f^{ij}\partial_j\phi^A)\approx
0 \ . \nonumber \\
\end{equation}
Note that  using (\ref{partiphimM}) we obtain another set of the
 primary constraints
\begin{eqnarray}\label{defSigmai}
\partial_i\phi^A\Pi_A=
%(p_A+ g_{mk}({}^3
%f^{-1})^{kl}\partial_l \phi_A \mR^m)=
%\partial_i\phi^A p_A+f_{il}({}^3 f^{-1})^{lk}
%g_{km}\mR^m=
\partial_i\phi^A
p_A+\mR_i=\Sigma_i\approx  0 \ .  \nonumber
\\
\end{eqnarray}
Observe that using (\ref{defSigmai}) we can write
\begin{eqnarray}
p_A+\mR_if^{ij}\partial_j\phi_A=\mM_{AC}\eta^{CB}p_B+\Sigma_i
f^{ij}\partial_j \phi_A
%\nonumber \\
%\Pi_A \eta^{AB}\Pi_B=p_D \eta^{DC}
%\mM_{CA}\eta^{AB}\mM_{BE}\eta^{EF}p_F+ 2\Sigma_i
%({}^{(3)}f^{-1})^{il}\partial_l \phi_A\eta^{AB}\mM_{BD}\eta^{DF}p_F+
%\nonumber \\
%+\Sigma_i ({}^{(3)}f^{-1})^{il}\partial_l \phi_A\eta^{AB}\partial_k
%\phi_B ({}^{(3)}f^{-1})^{km}\Sigma_m \approx
%\nonumber \\
%p_D \eta^{DC} \mM_{CA}\eta^{AF}p_F
%\nonumber \\
\end{eqnarray}
so that we can rewrite $\Sigma_p$ into the form
\begin{equation}
\Sigma_p= (\tn^i \mR_i+M_p^2\sqrt{g}U)^2+p_A\mM^{AB}p_B+H^i\Sigma_i
 \  \ ,
\end{equation}
where $H^i$ are  functions of the phase space variables. As a result
we see that it is natural to consider following independent
constraint $\Sigma_p$
\begin{equation}\label{defSigmap}
\Sigma_p= (\tn^i \mR_i+M_p^2\sqrt{g}U)^2+p_A\mM^{AB}p_B\approx 0 \ .
\end{equation}
We return to the analysis of the constraint $\Sigma_p$ below.

Now we are ready to write the extended Hamiltonian which includes
all the primary constraints
\begin{equation}
H_E=\int d^3\bx (N\mC_0+v_N\pi_N+v^i\pi_i+\Omega_p\Sigma_p+
\Omega^i\tSigma_i) \ ,
\end{equation}
where
\begin{eqnarray}
%H_0=\int d^3\bx [N(\frac{1}{\sqrt{g}M_p^2}
%\pi^{ij}\mG_{ijkl}\pi^{kl}-M_p^2 R -6m^2M_p^2\sqrt{g}) -\nonumber
%\\
%- (\tn^i\mR_i+M_p^2\mV)M-
% g_{ik}({}^3
%f^{-1})^{kl}f_{l0}\mR^i+M_p^2 \sqrt{g}M\mV+ +2m^2M_p^2\sqrt{g}N
%\sqrt{\tx}\tD^i_{ \ i} +
%\nonumber \\
% +(M\tn^i +({}^3 f^{-1})^{ik}f_{0k}+N
%\tD^i_{ \ j}\tn^j)\mR_i]= \nonumber \\
%=\int d^3\bx[N(\frac{1}{\sqrt{g}M_p^2}
%\pi^{ij}\mG_{ijkl}\pi^{kl}-M_p^2 R+ 2m^2\sqrt{g}\sqrt{\tx}\tD^i_{ \
%i} -6m^2M_p^2\sqrt{g})\equiv
%\nonumber \\
%=\int d^3\bx N\mC_0 \ , \quad
\mC_0= \frac{1}{\sqrt{g}M_p^2} \pi^{ij}\mG_{ijkl}\pi^{kl}-M_p^2
\sqrt{g} R+ 2m^2M_p^2\sqrt{g}\sqrt{\tx}\tD^i_{ \ i}
-6m^2M_p^2\sqrt{g}+
\tD^i_{ \ j}\tn^j\mR_i \nonumber \\
%=\int d^3\bx[N(\frac{1}{\sqrt{g}M_p^2}
%\pi^{ij}\mG_{ijkl}\pi^{kl}-M_p^2 {}^3R
%-6m^2M_p^2\sqrt{g})+\nonumber
%\\
%+\frac{1}{4(\tn^i\mR_i+M_p^2\sqrt{g}
%\sqrt{\tx}(1+N\tD^i_{ \ i}))}\times
%\nonumber \\
%\times  (p_A+ \mR_k ({}^3f^{-1})^{kl}
%\partial_l\phi_A)\eta^{AB}(p_B+\mR_m
%({}^3 f^{-1})^{mn}\partial_m\phi_B) +
% N
%\tD^i_{ \ j}\tn^j\mR_i] \nonumber \\
\end{eqnarray}
%using
%\begin{eqnarray}
%p_A\partial_0\phi^A
%%(\tn_i\mR^i+M_p^2\mV)\eta_{AC}\mM^C_{ \
%%B}\partial_t\phi^B\partial_t\phi^A-
%% g_{ik}({}^3
%%f^{-1})^{kl}f_{l0}\mR^i= \nonumber \\
%=- (\tn_i\mR^i+M_p^2\mV)M-
% g_{ik}({}^3
%f^{-1})^{kl}f_{l0}\mR^i \nonumber \\
%\end{eqnarray}
%Note also that the extended Hamiltonian has the form
and where we introduced the constraints $\tSigma_i$ defined as
\begin{equation}
\tSigma_i=\Sigma_i+\partial_i \tn^i\pi_i+
\partial_j(\tn^j\pi_i) \ .
\end{equation}
Note that $\tSigma_i$ is defined as    linear combination of the
constraints $\Sigma_i\approx 0$ together with  the constraints
$\pi_i\approx 0$.
% Observe that we have $4$ primary constraints
%$\pi_N\approx 0 \ , \pi_i\approx 0$, three primary constraints
%$\Sigma_i\approx 0$ and one primary constraint $\Sigma_p\approx 0$.
%However it turns out that it is useful to consider  $\tSigma_i$ in
%the form
%constraint $\tSigma_i$
%instead of $\Sigma$ which is defined as
%\begin{equation}
%\tSigma_i=\Sigma_i+ (\partial_i \tn_j -\partial_j \tn_i)\pi^j
%-\tn_i\partial_j\pi^j
%\end{equation}
%where it runs out it is more convenient to consider vector $\tn_i$
%rather then $\tn^i$.
%In fact, since $\tx$ depends on $\tn^i$ rather than $\tn_i$ it is

To proceed further we have to check the stability of all
constraints. To do this  we have to calculate the Poisson brackets
between all constraints and the Hamiltonian $H_E$. Note that  we
have following set of the canonical variables
$g_{ij},\pi^{ij},\phi^A,p_A,\tn^i,\pi_i$ and $N,\pi_N$ with
 non-zero Poisson brackets
\begin{eqnarray}\label{defpb}
\pb{g_{ij}(\bx),\pi^{kl}(\by)}&=& \frac{1}{2} (\delta_i^k\delta_j^l+
\delta_i^l\delta_j^k)\delta(\bx-\by) \ , \quad
\pb{\phi^A(\bx),p_B(\by)}=\delta^A_B\delta(\bx-\by) \ , \nonumber \\
\pb{N(\bx),\pi_N(\by)}&=&\delta(\bx-\by) \ , \quad
\pb{\tn^i(\bx),\pi_j(\by)}=\delta^i_j\delta(\bx-\by) \ . \nonumber
\\
\end{eqnarray}
Now we show that the smeared form of the constraint
$\tSigma_i\approx 0$
\begin{equation}\label{defbTS}
 \bT_S(\zeta^i)=\int d^3\bx
\zeta^i\tSigma_i \
\end{equation}
 is the generator of the spatial diffeomorphism.
 First of all using  (\ref{defpb}) we find
\begin{eqnarray}\label{pbbTSni}
\pb{\bT_S(\zeta^i),\tn^k}
%= \int d^3\by [ \zeta^i (-\partial_i
%\tn^j\delta_j^k\delta(\bx-\by))- \zeta^i\partial_j (\tn^j\delta_i^k
%\delta(\bx-\by))]= \nonumber \\
=-\zeta^i\partial_i\tn^k+\tn^j\partial_j\zeta^k \nonumber \\
\end{eqnarray}
which is  the correct transformation rule for $\tn^i$.
% Explicitly,
%by definition we have
%\begin{equation}
%\tn'^i(\bx')=\tn^j(\bx)\frac{\partial x'^i}{\partial x^j}
%\end{equation}
%and consequently
%\begin{equation}
% \tn'^i(\bx)-\tn^i(\bx) \equiv \delta
%\tn^i(\bx)=-\zeta^k(\bx)\partial_k\tn^i(\bx)+ \tn^k(\bx)\partial_k
%\zeta^i(\bx)
%\end{equation}
%which agrees with (\ref{pbbTSni}).
%\begin{eqnarray}
%\pb{\bT_S(N^i),\tn_k}= \int d^3\bx [- N^i(\partial_i\tn_j-\partial_j
%\tn_i)\delta^j_k\delta(\bx) +
%N^i \tn_i\partial_k (\delta(\bx))]= \nonumber \\
%=-N^i\partial_i \tn_k+ N^i\partial_k \tn_i -\partial_k (N^i\tn_i)=
%-N^i\partial_i \tn_k -\partial_k N^i\tn_i
%\nonumber \\
%\end{eqnarray}
%which is the correct transformation law since
%\begin{eqnarray}
%\tn_i(\bx')=\tn_j(\bx)\frac{\partial x^j}{
%\partial x'^i}
%\Rightarrow \nonumber \\
%\tn'_i(\bx) +\partial_j\tn_iN^j= \tn_j(\delta^j_i-
%\partial_i N^j)
%\Rightarrow \nonumber \\
%\tn'_i(\bx)-\tn_i(\bx)= \delta \tn_i= -\partial_j\tn_iN^j- \tn_j
%\partial_i N^j \ .
%\nonumber \\
%\end{eqnarray}
Then using (\ref{defpb})  we find
\begin{eqnarray}
\pb{\bT_S(N^i),\mR_j}&=&-\partial_i N^i \mR_j
-N^i\partial_i\mR_j-\mR_i\partial_j N^i \ ,
 \nonumber \\
\pb{\bT_S(N^i),p_A}&=&-N^i\partial_ip_A-
\partial_i N^ip_A \ , \nonumber \\
\pb{\bT_S(N^i),\phi^A}&=&-N^i\partial_i\phi^A \ ,
\nonumber \\
%\pb{\bT_S(N^i),\sqrt{g}}&=&-N^i\partial_i \sqrt{g}-\partial_i
%N^i\sqrt{g} \ .  \nonumber \\
\pb{\bT_S(N^i),g_{ij}}&=& -N^k\partial_k g_{ij}-\partial_i
N^k g_{kj}-g_{ik}\partial_j N^k \ , \nonumber \\
\pb{\bT_S(N^i),\pi^{ij}}&=& -\partial_k (N^k \pi^{ij}) +
\partial_k N^i\pi^{kj}+\pi^{ik}\partial_k N^j \ , \nonumber \\
%\nonumber \\
%\pb{\bT_S(N^i),\frac{1}{g}p_A p_B(\bx)}= -N^i\partial_i\left(
%\frac{1}{g}p_A p_B\right)(\bx) \ ,
%\nonumber \\
%\pb{\bT_S(N^i),\partial_i\phi^A}&=& -\partial_i N^k\partial_k\phi^A-
%N^k\partial_k (\partial_i\phi^A) \ ,
%\nonumber \\
\pb{\bT_S(N^i),f_{ij}}&=&-N^k\partial_k f_{ij}-\partial_i N^k
f_{kj}-f_{ik}
\partial_j N^k \ , \nonumber \\
%\pb{\bT_S(N^i), ({}^3 f^{-1})^{ij}}= -({}^3
%f^{-1})^{ik}\pb{\bT_S(N^i),f_{kl}} ({}^3 f^{-1})^{lj}=
%\nonumber \\
%-N^k \partial_k ({}^3 f^{-1})^{ij}+ ({}^3 f^{-1})^{ik}\partial_k
%N^j+\partial_k N^i ({}^3 f^{-1})^{kj}
%\nonumber \\
\pb{\bT_S(N^i),\pi^i}&=& -\partial_i N^i\pi^j -N^i\partial_i
\pi^j+\partial_j N^i\pi^j \ ,
 \nonumber \\
%\pb{\bT_S(N^i),\tx(\bx)}=-N^k\partial_k\tx(\bx) \nonumber \\
%\pb{\bT_S(N^i),(\partial_i \tn_j-\partial_j\tn_i)(\bx)}=
%-N^k\partial_k (\partial_i\tn_j-\partial_j\tn_i) -\partial_iN^m
%(\partial_m\tn_j-
%\partial_j\tn_m)-
%(\partial_i \tn_m-\partial_m\tn_i)\partial_jN^m
% \nonumber \\
% \pb{\bT_S(N^i),\Pi_A(\bx)}=-N^i(\bx)\partial_i\Pi_A(\bx)
% -\partial_iN^i(\bx)\Pi_A(\bx) \ , \nonumber \\
% \nonumber \\
% \pb{\bT_S(N^i),\Sigma_p}&=&-N^m\partial_m \Sigma_p
% -\partial_m N^m\Sigma_p \   \nonumber \\
 \end{eqnarray}
 that are the
  correct transformation rules of the canonical variables
under spatial diffeomorphism. To proceed further we need the Poisson
bracket between $\bT_S(N^i)$ and $\tD^i_{ \ j}$. It turns out that
it is convenient to know the explicit form of the matrix $\tD^i_{ \
j}$
\cite{Hassan:2011vm,Hassan:2011hr, Hassan:2011zd,Hassan:2011tf}
%\begin{equation}
%D^i_{ \ m}(\tx \delta^m_{ \ n} +\tn^m\tn^p f_{pn}) \tD^p_{ \ j}=
%g^{ik}f_{kj}=D^i_{ \ m}Q^m_{ \ p} \tD^p_{ \ j}= g^{ik}f_{kj}
%\end{equation}
%We multiply this equation from the right by $Q$ and take the square
%root so that we obtain
\begin{equation}\label{defDijagain}
\tD^i_{ \ j}= \sqrt{ g^{im}f_{mn}Q^n_{ \ p}} (Q^{-1})^p_{\ j} \ ,
\end{equation}
where
\begin{equation}\label{defQijagain}
Q^i_{ \ j}=\tx \delta^i_{ \ j} +\tn^i\tn^k f_{kj}
 \ , \quad
(Q^{-1})^p_{ \ q}= \frac{1}{\tx} (\delta^p_{ \ q} -\tn^p \tn^m
f_{mq})
 \ .
 \end{equation}
Using this expression we can easily determine the Poisson brackets
between $\bT_S(N^i)$ and $\tD^k_{ \ j}$. In fact, by definition we
have
\begin{eqnarray}
\pb{\bT_S(N^i),Q^k_{ \  l}}&=&-N^m\partial_m Q^k_{ \ l}+\partial_m
N^k
Q^m_{ \ l}-Q^k_{ \ n}\partial_l N^n \ , \nonumber \\
\pb{\bT_S(N^i),Q^i_{ \ j}Q^j_{ \ k}}&=& -N^m\partial_m( Q^i_{ \ j}
Q^j_{ \ k})+\partial_m N^i Q^i_{ \ j} Q^j_{ \ k}-Q^i_{ \ j}Q^j_{ \
m}\partial_k N^m \ . \nonumber \\
% \nonumber \\
%\pb{\bT_S(N^i), g^{im}f_{mn}Q^n_{ \ p}} =-N^q\partial_q
%(g^{im}f_{mn}Q^n_{ \ p}) +\partial_o N^i (g^{om}f_{mn}Q^n_{ \ p})
%-(g^{im}f_{mn}Q^n_{ \ o})\partial_p N^o\equiv \delta Q^k_{ \ l} \nonumber \\
\end{eqnarray}
%Observe that we have
%\begin{eqnarray}
%\delta (Q^i_{ \ j}Q^j_{ \ k})= \delta Q^i_{ \ j}Q^j_{ \ k}+Q^i_{ \
%j}\delta Q^j_{ \ k}=
%\nonumber \\
%(-N^m \partial_m Q^i_{ \ j}+\partial_m N^i Q^m_{ \ j} -Q^i_{ \
%m}\partial_j N^m)Q^j_{ \ k}+ Q^i_{ \ j}(-N^m\partial_m Q^j_{ \ k}
%+\partial_m N^j Q^m_{ \ k}-Q^j_{ \ m}\partial_i N^m)=
%\nonumber \\
%=-N^m\partial_m( Q^i_{ \ j} Q^j_{ \ k})+\partial_m N^i (Q^i_{ \ j}
%Q^j_{ \ k}-Q^i_{ \ j}Q^j_{ \ m}\partial_k N^m \ .
%\nonumber \\
%\end{eqnarray}
Using (\ref{defDijagain}) and the results derived above  we find
%\begin{eqnarray}
%\pb{\bT_S(N^i),g^{im}f_{mn}Q^n_{ \ j}}= -N^p\partial_p(
%g^{im}f_{mn}Q^n_{ \ j}) +\partial_p N^i (g^{pm}f_{mn}Q^n_{ \ j})-
%g^{im}f_{mn}Q^n_{ \ p}\partial_k N^p
% \ , \nonumber \\
% \end{eqnarray}
%As a result we find
\begin{equation}
\pb{\bT_S(N^i),\tD^i_{ \ j}}= -N^m\partial_m \tD^i_{ \ j}+
\partial_m N^i\tD^m_{ \ j}-
\tD^i_{ \ m}\partial_j N^m \ .
\end{equation}
Collecting all these results and after some calculations we find
\begin{eqnarray}\label{pbbTSmC0}
%\pb{\bT_S(N^i),V}=-N^m\partial_m V \ , \quad \pb{\bT_S(N^i),U}=
%-N^m\partial_m U \ , \nonumber \\
\pb{\bT_S(N^i),\mC_0}&=&-N^m\partial_m \mC_0-
\partial_m N^m \mC_0 \ , \nonumber \\
 \pb{\bT_S(N^i),\Sigma_p}&=&-N^m\partial_m \Sigma_p
 -\partial_m N^m\Sigma_p \ .  \nonumber \\
\end{eqnarray}
Note also that it is easy to show that following Poisson bracket
holds
\begin{equation}\label{bTSNM}
\pb{\bT_S(N^i),\bT_S(M^j)}= \bT_S(N^j\partial_j M^i-M^j
\partial_j N^i) \ .
\end{equation}
Now we are ready to analyze the stability of all
 primary
constraints. As usual the
 requirement of the preservation of the
constraint $\pi_N\approx 0$ implies an existence of the secondary
constraint $\mC_0\approx 0$. However the fact that $\mC_0$ is the
constraint immediately implies that  the constraint
$\tSigma_i\approx 0$ is preserved during the time evolution of the
system, using  (\ref{pbbTSmC0}) and
 (\ref{bTSNM}).
 % This result is also reflection that $\tSigma_i$
% are the constraints corresponding to the invariance of the theory
% under spatial diffeomorphism.
%%using also the fact that
%\begin{equation}
%\delta (Q^{-1})^i_{ \ j}= -(Q^{-1})^i_{ \ m}\delta Q^m_{ \ n}
%(Q^{-1})^n_{ \ j}=-N^m\partial_m (Q^{-1})^i_{ \ j} -(Q^{-1})^i_{ \
%k}\partial_j N^k+\partial_k N^i (Q^{-1})^k_{ \ j}
%\end{equation}
%As a result we find
%\begin{equation}
%\partial_t\tSigma_i(\bx)=
%\pb{H_T,\Sigma_i(\bx)}\approx 0 \ .
%\end{equation}
Now we analyze the requirement of the preservation of  the constraints
$\pi_i\approx 0$ during the time evolution of the system
\begin{eqnarray}
\partial_t\pi_i=\pb{\pi_i,H_E}=
%-N\frac{\partial (\tD^k_{ \ j}\tn^j)} {\partial \tn^i} (\mR_k
%-2m^2M_p^2\frac{\sqrt{g}}{\sqrt{\tx}} f_{kj}\tn^j)- \nonumber \\
%-\Omega_p \mR_i-\Omega_p  2m^2M_p^2 \frac{\sqrt{g}}
%{\sqrt{\tx}}f_{ij}\tn^j=\nonumber \\
%%=(\Omega_p g^{im}-
%% +
- \left(\Omega_p\delta^k_i+\frac{\partial (\tD^{k}_{ \ j}\tn^j)}
{\partial
\tn^i}\right)\left(\mR_k-2m^2M_p^2\frac{\sqrt{g}}{\sqrt{\tx}}f_{km}
\tn^m \right)= 0 \ .
\end{eqnarray}
It turns out that the following matrix
\begin{equation}
\Omega_p \delta^k_i +\frac{\partial (\tD^k_{ \ j}\tn^j)} {\partial
\tn^i}=0 \
\end{equation}
cannot be solved for $\Omega_p$ and hence we have to demand the
existence of  following secondary constraints
\cite{Hassan:2011vm,Hassan:2011hr, Hassan:2011zd,Hassan:2011tf}
\begin{equation}
\mC_i\equiv \mR_i-\frac{2m^2M_p^2\sqrt{g}}{\sqrt{\tx}} f_{ij} \tn^j
\approx 0 \ .
\end{equation}
%Now the requirement of the preservation of the constraint
%$\pi_N\approx 0$ gives
%\begin{equation}
%\partial_t \pi_N=\pb{H_E,\pi_N}=\mC_0\approx 0 \ .
%\end{equation}
%where $\mC_0\approx 0$ is the secondary constraint. Further, using
%the previous results we easily see that the constraints $\Sigma_i$
%are preserved during the time evolution of the system since the
%Hamiltonian $H_E$ is given as linear combinations of the constraints
%where these constraints have weakly vanishing Poisson brackets with
%$\bT_S(N^i)$.
%
Finally we have to proceed to the analysis of the time development
of the constraint $\Sigma_p\approx 0$. However it turns out that it
is very difficult to perform this analysis for $\Sigma_p$ due to the
presence of the terms that contain the spatial derivatives of
$\phi^A$. Then the explicit calculation gives
$\pb{\Sigma_p(\bx),\Sigma_p(\by)}\neq 0$.
 For that reason we proceed in a
 different way when we try  to simplify the constraint $\Sigma_p$.
   Using $\mC_i$ and $\Sigma_i$ we  find that
\begin{eqnarray}\label{defA}
%\mR_i f^{ij}\mR_j=\frac{4m^4 M_p^4 g \tn^i f_{ij}\tn^j}{\tx} +\mC_i
%f^{ij}\mC_j+\frac{4m^2M_p^2\sqrt{g}}{\sqrt{\tx}}\tn^i\mC_i\equiv
%\frac{4m^4 M_p^4 g\tn^i f_{ij}\tn^j}{\sqrt{\tx}}+ G^i \mC_i
%\Rightarrow
% \nonumber \\
% p_A\partial_i \phi^A f^{ij}
% \partial_j\phi^B p_B+\Sigma_i f^{ij}\Sigma_j-2\Sigma_i
% f^{ij}p_B\partial_j\phi^B=\frac{4m^4 g\tn^i f_{ij}\tn^j}{\sqrt{\tx}}+ G^i \mC_i
% \Rightarrow
% \nonumber \\
 A=\frac{4m^4 M_p^4 g\tn^i f_{ij}\tn^j}{\sqrt{\tx}}+ F^i\Sigma_i+G^i\mC_i
 \ ,  \nonumber \\
\end{eqnarray}
where
\begin{equation}
A=p_A\partial_i\phi^A f^{ij}
\partial_j\phi^Bp_B  \ ,
\end{equation}
and where $F^i,G^i$ are the phase space functions whose explicit
form is not important for us. Then with the help of (\ref{defA}) we
 express $\tn^if_{ij}\tn^j$ as a function of the  phase space variables
$p_A,\phi^A$ and $g_{ij},\pi^{ij}$
\begin{equation}\label{tni1}
\tn^i
f_{ij}\tn^j=\frac{A-F^i\Sigma_i-G^i\mC_i}{(A-F^i\Sigma_i-G^i\mC_i)+4m^4M_p^4
g} \ .
\end{equation}
In the same way we obtain
\begin{equation}\label{tni2}
\tn^i\mR_i=\frac{A-F^i\Sigma_i-G^i\mC_i}{\sqrt{(A-F^i
\Sigma_i-G^i\mC_i)+4m^4M_p^4g}}+\tn^i\mC_i \ ,
%\sqrt{\tx}=\frac{2m^2\sqrt{g}}{\sqrt{(A-F^i\Sigma_i-G^i)\mC_i+4m^4M_p^4g}}
%\ .
\end{equation}
and
\begin{equation}\label{tni3}
 \tn^i=-\frac{\partial_j\phi^Ap_A
f^{ji}}{\sqrt{A+4m^4M_p^4 g}}+ \tilde{F}^i\Sigma_i+\tilde{G}^i\mC_i
\ ,
\end{equation}
where again $\tilde{F}^i,\tilde{G}^i$ are phase space functions
whose explicit form is not needed for us.

Now using these results we find that the  constraint
 $\Sigma_p$ takes the form
\begin{eqnarray}
\Sigma_p&=&
%\tn^i\mR_i\tn^j\mR_j+4m^2M_p^2
%\sqrt{g}\tn^i\mR_i\sqrt{\tx}+4m^4M_p^4 g\tx+ p_Ap^A-A=
%\nonumber \\
\frac{(A-F^i\Sigma_i-G^i\mC_i+4m^4 M_p^4 g)4m^4 M_p^4
g}{A-F^i\Sigma_i-G^i\mC_i+4m^4 M_p^4 g}+H^i\Sigma_i+p_A
p^A=\nonumber \\
&=& p_A
p^A+4m^4 M_p^4 g+H^i\Sigma_i \equiv 4m^4M_p^4 g\tSigma_p+H^i\Sigma_i \ , \nonumber \\
\end{eqnarray}
where we introduced new independent constraint $\tSigma_p$
\begin{equation}\label{Sigmapfin}
\tSigma_p=\frac{p_A p^A}{4m^4 M_p^4 g}+1=0 \
\end{equation}
that has precisely the same form as in \cite{Kluson:2012wf}. Note
that the constraint $\tSigma_p$ has the desired  property that
$\pb{\tSigma_p(\bx),\tSigma_p(\by)}=0$. As a result we see that it
is more natural  to consider $\tSigma_p$ instead of $\Sigma_p$ as an
independent constraint. Then the total Hamiltonian, where we include
all constraints, takes the form
\begin{equation}\label{HTfin}
H_T=\int d^3\bx (N\mC_0+v_N\pi_N+v^i\pi_i+ \Omega_p \tSigma_p+
\Omega^i\tSigma_i+\Gamma^i\mC_i) \ .
\end{equation}
 Now we are ready to
analyze the stability of all constraints that appear in
(\ref{HTfin}).  First of all we find that $\pi_N\approx 0 $ is
automatically preserved while the preservation of the constraint
$\pi_i\approx 0$ gives
\begin{eqnarray}\label{parttpi}
\partial_t\pi_i&=&\pb{\pi_i,H_T}\approx
\int d^3\bx \Gamma^j(\bx)\pb{\pi_i,\mC_j(\bx)} = \nonumber \\
&=&-2m^2\Gamma^j\frac{1}{\sqrt{\tx}} (f_{ij}-f_{ik}\tn^k
f_{jl}\tn^l)\equiv -\triangle_{\pi_i,\mC_j}\Gamma^j \ .
\nonumber \\
\end{eqnarray}
By definition
\begin{eqnarray}
\det (f_{ij}-f_{ik}\tn^k f_{il}\tn^l)
%= \det f_{ij} \det (\delta^j_k
%-f^{jm}f_{mn}\tn^n f_{kl}\tn^l)= \nonumber \\
%\det f_{ij}\exp \ln (1-\tn^i f_{ij}\tn^j)=
=\tx \det f_{ij}\neq 0
 \nonumber \\
\end{eqnarray}
and hence the matrix $\triangle_{\pi_i,\mC_j}$  is non-singular.
Then the only solution of the equation (\ref{parttpi}) is
 $\Gamma^i=0$.

As the next step   we proceed to the analysis of the stability of
the constraint $\tSigma_p$. As is clear from (\ref{Sigmapfin}) we
have
\begin{equation}
\pb{\tSigma_p(\bx),\tSigma_p(\by)}=0 \ .
\end{equation}
Then the time evolution of given constraint takes the form
\begin{eqnarray}\label{timeSigmap}
\partial_t \tSigma_p=\pb{\tSigma_p,H_T}\approx
\int d^3\bx N(\bx)\pb{\Sigma_p,\mC_0(\bx)} \
\nonumber \\
\end{eqnarray}
using the fact that $\tSigma_p$ does not depend on $\tn^i$ together
with  $\Gamma^i=0$ and also the fact that $\tSigma_p$ is manifestly
diffeomorphism invariant.

In order to explicitly determine (\ref{timeSigmap})
 we need
following expression
\begin{eqnarray}
\frac{\delta ( \sqrt{\tx}\tD^k_{ \ k})}{\delta f_{ij}}
%=\frac{1}{2} \tr
%(\bA^{-1/2})^k_{ \ l}\frac{\delta \bA^l_{ \ k}}{\delta f_{ij}}=
%\nonumber \\
%= \frac{1}{2\sqrt{\tx}}(\tD^{-1})^j_{ \
%l}((g^{li}-\tD^l_m\tn^m\tD^i_{ \ n}\tn^n)-\frac{1}{\sqrt{\tx}} \tn^l
%f_{lm}\frac{\delta \tD^l_{ \ p}\tn^p}{\delta f_{ij}}= \nonumber
%\\
%\frac{\sqrt{\tx}}{2}(\tD^{-1})^j_{ \ k}\tD^k_{ \ m}\tD^m_{ \ p}
%f^{pi} -\frac{1}{\sqrt{\tx}} \tn^l f_{lm}\frac{\delta \tD^l_{ \
%p}\tn^p}{\delta f_{ij}} =\nonumber \\
=\frac{\sqrt{\tx}}{2}\tD^j_{ \ p} f^{pi} -\frac{1}{\sqrt{\tx}} \tn^l
f_{lm}\frac{\delta( \tD^m_{ \
p}\tn^p)}{\delta f_{ij}} \ . \nonumber \\
\end{eqnarray}
%where $\bA^i_{ \ j}=(g^{im}-\tD^i_{ \ n}\tn^n \tD^m_{ \
%l}\tn^l)f_{mj}$
 Then after some calculations we obtain
%Then we have
%\begin{eqnarray}
%\pb{\Sigma_p,\int d^3\bx N \tD^i_{ \ j} \tn^j \mR_i(\bx)}\approx
%-\pb{\Sigma_p,\int d^3\bx N\tD^i_{ \ j}\tn^j\partial_i\phi^A
%p_A(\bx)}= \nonumber \\
%=-\frac{p_A}{2M_p^4m^4 g}\partial_i[N\tD^i_{ \ j}\tn^j p^A]
%-N\frac{4p_A}{4M_p^4m^4 g}\partial_i[N\frac{\delta (\tD^k_{ \
%l}\tn^l)}{\delta f_{ij}}\partial_k\phi^C p_C\partial_j\phi^A]
%\nonumber \\
%\end{eqnarray}
%and
%\begin{eqnarray}
%\pb{\Sigma_p,2m^2M_p^2\int d^3 \bx \sqrt{\tx}\sqrt{g}\tD^i_{ \ i}}=
%-2m^2\frac{4p_A}{4M_p^4 m^4g}\int d^3\bx N\sqrt{g}\frac{\delta
%(\sqrt{\tx}\tD^k_{ \ k})} {\delta
%f_{ij}}\partial_i\delta(\bx)\partial_j\phi^A(\bx)=
%\nonumber \\
%=2m^2M_p^2\frac{4p_A}{4M_p^4 m^4 g}\partial_i[N\sqrt{g}\frac{\delta
%(\sqrt{\tx}\tD^k_{ \ k})}
%{\delta f_{ij}}\partial_j\phi^A]=\nonumber \\
%=2m^2M_p^2\frac{4p_A}{4M_p^4 m^4 g}\partial_i[N\sqrt{g}
%\frac{\sqrt{\tx}}{2}\tD^j_{ \ p} f^{pi}\partial_j\phi^A
%-\frac{\sqrt{g}}{\sqrt{\tx}} \tn^l f_{lm}\frac{\delta \tD^m_{ \
%p}\tn^p}{\delta f_{ij}}\partial_j\phi^A
%] \nonumber \\
%\end{eqnarray}
%so that we obtain
\begin{eqnarray}
& &\pb{\tSigma_p,\int d^3\bx N\mC_0}=2 \partial_i[N\tD^i_{ \ j}]\tn^j\Sigma_p + \nonumber \\
&+&\frac{1}{M_p^4m^4 g}p_A\partial_i\left[N\frac{\delta (\tD^k_{ \
l} \tn^l)}{\delta f_{ij}} \mC_k \partial_j\phi^A\right]
-\frac{1}{M_p^4m^4
g}p_A\partial_j\phi^A\partial_i\left[N\frac{\delta (\tD^k_{ \ l}
\tn^l)}{\delta f_{ij}} \right]\Sigma_k+\nonumber \\
&+&N\left(-\tD^i_{ \ j}\frac{\partial_i[\tn^j p_A]p_A}{2m_p^4 m^4 g}
+\frac{2m^2 M_p^2}{M_p^4m^4 g}\tD^i_{ \ k}p_A
\partial_i[\sqrt{g}\sqrt{\tx}f^{kj}\partial_j \phi^A]
\right)\approx
N\Sigma_p^{II} \ .  \nonumber \\
\end{eqnarray}
%using also
%\begin{eqnarray}
%\partial_i[N\tD^i_{ \ p}](-
%\frac{p_A}{2M_p^4m^4 g}\tn^p p^A+
%\frac{m^2M_p^2\sqrt{g}\sqrt{\tx}}{M_p^4m^4 g}
%f^{pj}\partial_j\phi^Ap_A)\approx \nonumber \\
%\partial_i[N\tD^i_{ \ p}](-
%\frac{p_A}{2M_p^4m^4 g}\tn^p p^A- \frac{m^2M_p^2\sqrt{g}}{M_p^4m^4
%g}
%f^{pj}\mR_j)=\nonumber \\
%\partial_i[N\tD^i_{ \ p}](-
%\frac{p_A}{2M_p^4m^4 g}\tn^p p^A- \frac{m^2M_p^2\sqrt{g}}{M_p^4m^4
%g}
%f^{pj}(\mC_j+\frac{2m^2M_p^2\sqrt{g}}{\sqrt{\tx}}f_{jn}\tn^n))\approx\nonumber \\
%\approx -2\partial_i[N\tD^i_{ \ p}](\frac{p_A p^A}{4M_p^4 m^4
%g}+1)\tn^p=-2 \partial_i[N\tD^i_{ \ p}]\tn^p\Sigma_p \nonumber \\
%\end{eqnarray}
%using also
%\begin{equation}
%\tD^i_{ \ k}f^{kj}=\tD^j_{ \ k}f^{ ki} \ .
%\end{equation}
In order to simplify $\Sigma_p^{II}$ further we use (\ref{tni1}),
(\ref{tni2}) together with (\ref{tni3}) to make $\Sigma_p^{II}$
independent on $\tn^i$. In fact, using
(\ref{defDijagain})(\ref{defQijagain}) we obtain
%  note that $\tD^i_{ \ j}$ has an
%explicit form
% To proceed
%further it turns out that it is useful to replace $\tn^i$ with
%\begin{equation} \tn^i=-\frac{\partial_j\phi^Ap_A
%f^{ji}}{\sqrt{A+4m^4M_p^4 g}}+ \tilde{F}^i\Sigma_i+\tilde{G}^i\mC_i
%\ .
%\end{equation}
%Using also
%\begin{equation}
%\tn^i
%f_{ij}\tn^j=\frac{A-F^i\Sigma_i-G^i\mC_i}{(A-F^i\Sigma_i-G^i\mC_i)+4m^4M_p^4
%g} \ , A=p_A\partial_i\phi^A f^{ij}
%\partial_j\phi^Bp_B \ .
%\end{equation}
%and then
%\begin{equation}
%\tn^i\mR_i=\frac{A-F^i\Sigma_i-G^i\mC_i}{\sqrt{(A-F^i
%\Sigma_i-G^i\mC_i)+4m^4M_p^4g}}+n^i\mC_i \ ,
%\sqrt{\tx}=\frac{2m^2\sqrt{g}}{\sqrt{(A-F^i\Sigma_i-G^i)\mC_i+4m^4M_p^4g}}
%\ .
%\end{equation}
%Then note that we can write $\tn^i$ as
%\begin{equation}
%\tn^i=-\frac{\partial_j\phi^Ap_A f^{ji}}{\sqrt{A+4m^4M_p^4 g}}+
%\tilde{F}^i\Sigma_i+\tilde{G}^i\mC_i \ .
%\end{equation}
%\begin{eqnarray}
%\tD^i_{ \ j}&=&\sqrt{g^{im}f_{mn}Q^n_{ \ p}}(Q^{-1})^p_{ \ j} \ ,
%\nonumber \\
% Q^m_{ \ p}&=&\tx \delta^m_{ \ p}+\tn^m\tn^n f_{np} \ ,
%(Q^{-1})^p_{ \ j}=\frac{1}{\tx}(\delta^m_p-\tn^p\tn^m f_{mj}) \ .
%\nonumber \\
%\end{eqnarray}
%and hence we find up to expressions proportional to $\Sigma_i,\mC_i$
\begin{eqnarray}
Q^m_p&=&\frac{1}{A+4M_p^4 m^4 g}(4m^4M_p^4g\delta^m_{ \ p}+
\partial_j\phi^Ap_A f^{jm}\partial_p \phi^Bp_B) \ , \nonumber \\
(Q^{-1})^m_p&=&\frac{A+4M_p^4 m^4 g}{4m^4 M_p^4 g}(\delta^m_{ \
p}-\frac{1}{A+4m^4 M_p^4 g}
\partial_j\phi^Ap_A f^{jm}\partial_p \phi^Bp_B) \  \nonumber \\
\end{eqnarray}
up to terms proportional to the constraints $\mC_i,\Sigma_i$.  With
the help of these results we obtain
\begin{eqnarray}
\Sigma_p^{II}&=& -\tD^i_{ \ j}\frac{\partial_i[\tn^j p_A]p_A}{2m_p^4
m^4 g} +\frac{2m^2 M_p^2}{M_p^4m^4 g}\tD^i_{ \ p}p_A
\partial_i[\sqrt{g}\sqrt{\tx}f^{pj}\partial_j \phi^A]+F'^i\Sigma_i+
G'^i\mC_i \equiv \nonumber \\
&\equiv & \tSigma_p^{II}+F'^i\Sigma_i+G'^i\mC_i \ , \nonumber \\
\end{eqnarray}
where $\tx$ and $\tD^i_{ \ j}$ are functions of
$p_A,\partial_j\phi^A$ and $g$ through the relations
(\ref{tni1}),(\ref{tni2}) and (\ref{tni3}).
%It is important that now $\Sigma_p^{II}$ does not depend on $\tn^i$
%and hence the Poisson bracket between $\Sigma_p^{II}$ and $\pi_i$ is
%zero.
Now we see from (\ref{timeSigmap}) that  the time evolution of the
constraint $\tSigma_p\approx 0$ is obeyed on condition when either
$N=0$ or when $\tSigma^{II}_p=0$. Note that we should interpreted
$N$ as the Lagrange multiplier so that it is possible to demand that
$N=0$ on condition when $\tSigma^{II}_p\neq 0$ on the whole phase
space. Of course such a condition is too strong so that it is more
natural to demand that $\tSigma_p^{II}\approx 0 $ and $N\neq 0$. In
other words $\tSigma_p^{II}\approx 0$ is the new   secondary
constraint.

In summary we have following collection of constraints:
$\pi_N\approx 0 \ , \pi_i\approx 0 \ ,  \mC_0\approx 0, \mC_i\approx
0, \tSigma_i\approx 0 , \tSigma_p\approx 0 , \tSigma^{II}_p\approx
0$. The dynamics of these constraints is governed by the total
Hamiltonian
\begin{equation}
H_T=\int d^3\bx (N\mC_0+v_N\pi_N+v^i\pi_i+\Omega_p\tSigma_p+
\Omega_p^{II}\tSigma_p^{II}+\Omega^i\tSigma_i+\Gamma^i\mC_i) \ .
\end{equation}
As the final step we have to analyze  the preservation of all
constraints. The case of $\pi_N\approx 0$ is trivial. For
$\pi_i\approx 0$ we obtain
\begin{eqnarray}\label{parttpii2}
\partial_i\pi_i(\bx)&=&\pb{\pi_i(\bx),H_T}=
\int d^3\by( \Gamma^j(\by)\pb{\pi_i(\bx),\mC_j(\by)}
+\Omega_p^{II}(\by) \pb{\pi_i(\bx),\tSigma_p^{II}(\by)})=
 \ \nonumber \\
&=& \Gamma^j\triangle_{\pi_i,\mC_j}(\bx)=0
%+\Omega_p^{II}(\by)\triangle_{\pi_i,\Sigma_p^{II}}(\bx,\by)) =0
\nonumber \\
\end{eqnarray}
due to the crucial fact that $\tSigma_p^{II}$ does not depend on
$\tn^i$. This is the main reason why we introduced $\tSigma_p^{II}$
instead of $\Sigma_p^{II}$. Then as we argued above the only
solution of the equation  is $\Gamma^i=0$.  Now the time development
of $\mC_i$ is given by the equation
\begin{eqnarray}\label{parttCi}
\partial_t \mC_i(\bx)&=&\pb{\mC_i(\bx),H_T}
\approx \nonumber \\
&\approx & \int d^3\bx \left(N(\by)\pb{\mC_i(\bx),\mC_0(\by)}+
v^j(\by)
\pb{\mC_i(\bx),\pi_j(\by)}+\right.\nonumber \\
&+&\left.\Omega_p(\by)\pb{\mC_i(\bx),\tSigma_p(\by)}+
\Omega_p^{II}(\by)
\pb{\mC_i(\bx),\tSigma_p^{II}(\by)}\right) \nonumber \\
\end{eqnarray}
and the time development of the constraint $\tSigma_p$ is governed
by the equation
\begin{eqnarray}
\partial_t\tSigma_p(\bx)=
\pb{\tSigma_p(\bx),H_T}\approx \int d^3\bx \Omega_p^{II}(\by)
\pb{\tSigma_p(\bx),\tSigma^{II}_p(\by)} \ .
\nonumber \\
\end{eqnarray}
As follows from the explicit form of the constraint $\tSigma_p^{II}$
we  see that $\pb{\tSigma_p^{II}(\bx),\tSigma_p(\by)}$ is non-zero
and proportional also to the  higher order  derivatives of the delta
functions.  As a consequence we find that the only solution of the
equation above is $\Omega_p^{II}=0$. Further we analyze the time
evolution of the constraint $\tSigma_p^{II}$
\begin{eqnarray}
\partial_t\tSigma_p^{II}(\bx)&=&\pb{\tSigma_p^{II}(\bx),H_T}=
\nonumber \\
&=&\int d^3\bx \left(N(\by)\pb{\tSigma_p^{II}(\bx),\mC_0(\by)}+
\Omega_p(\by) \pb{\tSigma^{II}_p(\bx),\tSigma_p(\by)}
\right)=0 \ .  \nonumber \\
\end{eqnarray}
 Now
from the last equation we obtain $\Omega_p$ as a function of the
phase space variables and $N$, at least in principle. Then inserting
this result into the equation for the preservation of $\mC_i$
(\ref{parttCi}) we determine $v^j$ as functions of the phase space
variables.
 Finally note also that the
constraint $\mC_0$ is automatically preserved due to the fact that
$\Gamma^i=\Omega_p^{II}=0$ and also the fact that
$\pb{\mC_0(\bx),\mC_0(\by)}\approx 0$ as was shown in
\cite{Hassan:2011ea}.

 In
summary we obtain following picture. We have five the first class
constraints $\pi_N\approx 0\ , \mC_0\approx 0 \ , \tSigma_i\approx
0$ together with eight the second class constraints $\pi_i\approx 0
\ , \mC_i\approx 0$ and $\tSigma_p\approx 0 \ ,
\tSigma_p^{II}\approx 0$. The constraints $\pi_i\approx 0$ together
with $\mC_i\approx 0$ can be solved for $\pi_i$ and $\tn^i$. Then
the constraint $\tSigma_p$ can be solved for one of the four momenta
$p_A$ while the constraint $\tSigma_p^{II}$ can be solved for one of
the four $\phi^A$. As a result we have $12$ gravitational degrees of
freedom $g_{ij},\pi^{ij}$,$6$ scalars degrees of freedom together
with $4$  first class constraints $\mC_0\approx 0 \ ,
\tSigma_i\approx 0$. Then we find that the number of physical
degrees of freedom is $10$ which is the correct number of physical
degrees of freedom of the massive gravity.

%\begin{eqnarray}
%\partial_t \mC_0(\bx)=
%\pb{\mC_0(\bx),H_T}\approx \nonumber \\
% \int d^3\by
%(\Omega_p^{II}(\by) \pb{\mC_0(\bx),
%\Sigma_p^{II}(\by)}+\Gamma^j(\by)\pb{\mC_0(\bx),\mC_j(\by)})=0
% \nonumber \\
%\end{eqnarray}

\section{General Non-Linear Massive Gravity Action}
Let us try to apply the procedure performed in previous section to
the case of the  general non-linear massive gravity whose  action
takes the form
\begin{equation}
S=M_p^2\int d^3\bx dt [N\sqrt{g}
\tK_{ij}\mG^{ijkl}\tK_{kl}+\sqrt{g}N R +2m^2 \sqrt{g}MU + 2m^2N
\sqrt{g} V] \ ,
\end{equation}
where
\begin{eqnarray}
U&=&\beta_1 \sqrt{\tx}+\beta_2[(\sqrt{\tx})^2 \tD^i_{ \ i}+\tn^i
f_{ij} \tD^j_{ \ k}\tn^k]+\nonumber \\
&+&\beta_3[\sqrt{\tx}(\tD^l_{ \ l}\tn^i f_{ij} \tD^j_{ \ k} \tn^k-
\tD^i_{ \ k}\tn^k f_{ij}\tD^j_l \tn^l)+ \frac{1}{2} \sqrt{\tx}^3
(\tD^i_{ \ i}\tD^j_{ \ j}- \tD^i_{ \ j }\tD^j_{ \ i})] \ ,  \nonumber \\
V&=&\beta_0+\beta_1\sqrt{\tx}\tD^i_{ \ i}+ \frac{1}{2} \sqrt{\tx}^2
[\tD^i_{ \ i}\tD^j_{ \ j}+ \tD^i_{ \ j}\tD^j_{ \ i}]+\nonumber \\
&+&\frac{1}{6}\beta_3\sqrt{\tx}^3 [\tD^i_{ \ i}\tD^j_{ \ j} \tD^k_{
\ k} -3\tD^i_{ \ i} \tD^j_{ \ k}\tD^k_{ \ j}+ 2\tD^i_{ \ j} \tD^j_{
\
k}\tD^k_{ \ i}] \nonumber \\
\end{eqnarray}
%Now the Hamiltonian analysis proceed as in the previous section with
%the bare Hamiltonian in the form
%\begin{eqnarray}
%H_0 =\int d^3\bx N\mC_0 \ , \quad \nonumber \\
%\mC_0= (\frac{1}{\sqrt{g}M_p^2} \pi^{ij}\mG_{ijkl}\pi^{kl}-M_p^2
%{}^{(3)}R -2m^2 M_p^2 V +\mR_i\tD^i_{ \ j} \tn^j) \ , \nonumber \\
%%=\int d^3\bx[N(\frac{1}{\sqrt{g}M_p^2}
%%\pi^{ij}\mG_{ijkl}\pi^{kl}-M_p^2 {}^3R
%%-6m^2M_p^2\sqrt{g})+\nonumber
%%\\
%%+\frac{1}{4(\tn^i\mR_i+M_p^2\sqrt{g}
%%\sqrt{\tx}(1+N\tD^i_{ \ i}))}\times
%%\nonumber \\
%%\times  (p_A+ \mR_k ({}^3f^{-1})^{kl}
%%\partial_l\phi_A)\eta^{AB}(p_B+\mR_m
%%({}^3 f^{-1})^{mn}\partial_m\phi_B) +
%% N
%%\tD^i_{ \ j}\tn^j\mR_i] \nonumber \\
%\end{eqnarray}
%using
%\begin{eqnarray}
%p_A\partial_0\phi^A
%%(\tn_i\mR^i+M_p^2\mV)\eta_{AC}\mM^C_{ \
%%B}\partial_t\phi^B\partial_t\phi^A-
%% g_{ik}({}^3
%%f^{-1})^{kl}f_{l0}\mR^i= \nonumber \\
%=- (\tn_i\mR^i+M_p^2\mV)M-
% g_{ik}({}^3
%f^{-1})^{kl}f_{l0}\mR^i \nonumber \\
%\end{eqnarray}
Following the analysis performed in the previous section we find the
extended Hamiltonian  in the form
\begin{equation}
H_E=\int d^3\bx (N\mC_0+v_N\pi_N+v^i\pi_i+\Omega_p\Sigma_p+
\Omega^i\tSigma_i) \ ,
\end{equation}
where
\begin{equation}
\mC_0= \frac{1}{\sqrt{g}M_p^2} \pi^{ij}\mG_{ijkl}\pi^{kl}-M_p^2
\sqrt{g}R -2m^2 M_p^2 V +\mR_i\tD^i_{ \ j} \tn^j \ ,
\end{equation}
and where
 the primary constraint $\Sigma_p$ takes the form
\begin{equation}\label{Sigmapgen}
\Sigma_p:(\tn^i \mR_i+2m^2M_p^2\sqrt{g}U)^2 +p_A\mM^{AB}p_B\approx 0
\ .
\end{equation}
As the next step we should analyze the stability of all primary
constraints. As in previous section we find that the stability of
the constraint $\pi_N\approx 0$ implies the secondary constraint
$\mC_0\approx 0$ while the stability of the constraints $\pi_i$
implies set of the secondary constraints $\mC_i$
\cite{Hassan:2011vm,Hassan:2011hr,Hassan:2011zd,Hassan:2011tf}
\begin{eqnarray}
\mC_i=\mR_i-2m^2\sqrt{g} \frac{\tn^l f_{lj}} {\sqrt{\tx}} \left[
\beta_1\delta_i^j+\beta_2\sqrt{\tx}(\delta_i^j\tD^m_{ \ m} -\tD^j_{
\ i})+ \right.
\nonumber \\
\left.+\beta_3\sqrt{\tx}^2\left(\frac{1}{2}\delta^j_{ \ i} (\tD^m_{
\ m}\tD^n_{ \ n}-\tD^m_{ \ n}\tD^n_{ \ m})+ \tD^j_{ \ m}\tD^m_{ \ i}
-\tD^j_{ \ i}\tD^m_{ \ m}\right)\right] \ . \nonumber \\
\end{eqnarray}
%In principle we can proceed as in previous section. In summary we
%have following collection of the primary constraints
%$\tSigma_i\approx 0 , \pi_i\approx 0 \ , \pi_N \approx 0,
%\Sigma_p\approx 0$ and we should analyze their stability. We again
%find that $\tSigma_i$ are preserved while $\pi_N\approx 0$ implies
%an existence of the constraint $\mC_0\approx 0$. Considering the
%constraints $\pi_i\approx 0$, we can follow
%\cite{Hassan:2011zd,Hassan:2011tf,Hassan:2011hr,Hassan:2011vm} and
%find that requirement of the preservation of the constraints
%$\pi_i\approx 0$ implies the secondary constraints
%
Now we come to the key point of the analysis which is the
requirement of the preservation of the constraint $\Sigma_p\approx
0$ during the time evolution of the system. This is very complicated
expression which depends on the all phase space variables. Remember
that in the minimal case we expressed  $\tn^i$ as  functions of
$g_{ij},p_A$ and $\phi^A$. As a result we found that $\Sigma_p$ can
be expressed  as a linear combination of $\mC_i,\Sigma_i$ and
$\tSigma_p$ where $\tSigma_p$ obeys an important property
$\pb{\tSigma_p(\bx),\tSigma_p(\by)}=0$.%
%Note however that clearly $\pb{\Sigma_p(\bx),\Sigma_p(\by)}\neq 0$
%due to the fact that $\pb{\tSigma_p(\bx),\mC_i(\by)}\neq 0$.
It would be certainly nice to repeat the same procedure in the case
of the constraint $\Sigma_p$ given in (\ref{Sigmapgen}).
Unfortunately we are not able to solve the constraint $\mC_i$ in
order to express $\tn^i$ as a function of $\mR_i$. Consequently we
are not able  to express the constraint  $\Sigma_p$ as a linear
combination of the constraints $\mC_i$ and possibly
$\tSigma_i,\mC_0$ and the new constraint $\tSigma_p$ where
$\pb{\tSigma_p(\bx),\tSigma_p(\by)}=0$. In other words, despite of
the fact we are able to find the primary constraint $\Sigma_p\approx
0$ we are not able to determine the additional secondary constraint
which is necessary for the elimination of two non-physical phase
space modes. As a result the main goal of this paper which was the
proof of the absence of the ghosts in the general non-linear massive
gravity action in St\"{u}ckelberg formalism cannot be completed.

 \noindent {\bf
Acknowledgement:}
 This work   was
supported by the Grant agency of the Czech republic under the grant
P201/12/G028. \vskip 5mm

\end{document}